\documentclass[print,showpacs,oneside,twocolumn,prl,amsmath,amssymb]{revtex4}
\usepackage{pifont}
\usepackage{cases}
\usepackage{amsmath}
\usepackage{amssymb}
\usepackage{amsfonts}
\usepackage{amssymb}
\usepackage{dcolumn}
\usepackage{bm}
\usepackage{graphicx}

\begin{document}

\title{NMR Experimental Demonstration of Probabilistic Quantum Cloning}
\author{Hongwei Chen$^{1,2}$}
\author{Dawei Lu$^{1}$}
\author{Bo Chong$^{1,3}$}
\author{Gan Qin$^{1}$}
\author{Xianyi Zhou$^{1}$}
\author{Xinhua Peng$^{1}$}
\email{xhpeng@ustc.edu.cn}
\author{Jiangfeng Du$^{1}$}
\email{djf@ustc.edu.cn}
\affiliation{$^1$Hefei National Laboratory for Physical Sciences at
Microscale and Department of Modern Physics, University of Science
and Technology of China, Hefei, Anhui 230026, People's Republic of
China.\\
$^2$High Magnetic Field Laboratory, Hefei Institutes of Physical Science,
Chinese Academy of Sciences, Hefei 230031, People's Republic of China\\
$^3$College of Science, Xi'an University of Architecture and Technology, Xi'an 710055, P. R. China}
\date{\today}

\begin{abstract}
The method of quantum cloning is divided into two main categories:
approximate and probabilistic quantum cloning. The former method is
used to approximate an unknown quantum state deterministically, and
the latter can be used to faithfully copy the state
probabilistically. So far, many approximate cloning machines have
been experimentally demonstrated, but probabilistic cloning remains
an experimental challenge, as it requires more complicated networks
and a higher level of precision control. In this work, we designed
an efficient quantum network with a limited amount of resources, and
performed the first experimental demonstration of probabilistic
quantum cloning in an NMR quantum computer. In our experiment, the
optimal cloning efficiency proposed by Duan  and Guo
[Phys.~Rev.~Lett.~\textbf{80}, 4999 (1998)] is achieved.

\end{abstract}

 \maketitle

The no-cloning theorem states that an arbitrary quantum state cannot
be cloned perfectly \cite{Wootters299}. This is a direct consequence of the linearity of quantum mechanics, and constitutes one of the most fundamental differences between the classical and quantum information theory. Remarkably, this property is also the crucial element for guaranteeing the security of many quantum key distribution protocols. Although quantum states can not be cloned faithfully, in a seminal paper,
Bu$\breve{\texttt{z}}$ek and Hillery proposed an "approximate
cloning machine" that can produce two identical copies approximately close to
the original one \cite{Buzek54}. On the other hand, quantum cloning machines are of
significant importance in quantum cryptography as they provide the
optimal eavesdropping technique for a large class of attacks on many
quantum key distribution protocols \cite{Bennett84,Niu60} and have
attracted a great deal of interest in further research
\cite{Scarani77,Gisin79,Cerf85,Bruss62,Ariano64,Delgado98,Lamata101}. Up to now, several
approximate cloning machines have been experimentally demonstrated
in optical systems
\cite{Huang64,Antia296,Martini419,Fasel89,Zhao95,Bartuskova99,Nagli720}
and NMR systems \cite{Cummins88,Du94,Chen75}.

Apart from the idea of an approximate cloning machine, an interesting
alternative quantum cloning machine, probabilistic quantum
cloning machine (PQCM), was proposed by Duan and Guo \cite{Duan80}.
They showed that states randomly chosen from a known set of states can be
probabilistically cloned with perfect fidelity, if the states in the set are linearly
independent. More specifically, if the
states in the set are two non-orthogonal states $|\psi_1\rangle$ and
$|\psi_2\rangle$, the \emph{cloning fidelity} will be 1 and the
optimal \emph{cloning efficiency} $\gamma$ is given by $
1/(1+|\langle\psi_1|\psi_2\rangle|)$.

There are many theoretical investigations on PQCM in
literatures \cite{Pati83,Qiu39,Azuma72}, and resource demanding experimental proposals exist \cite{Zhang61}. However, to the best of our knowledge, no experiment has been reported until now.
To make the experiment feasible, one should overcome two
difficulties: (i) to minimize the quantum network complexity, and
(ii) to achieve precise quantum control at a certain error
threshold.  In this work,  we successfully solved these two
problems and experimentally demonstrated the probabilistic quantum
cloning machine with optimal cloning efficiency \cite{Duan80}.

\begin{figure}[tbph]
  \centering
  \includegraphics[width=8.5 cm]{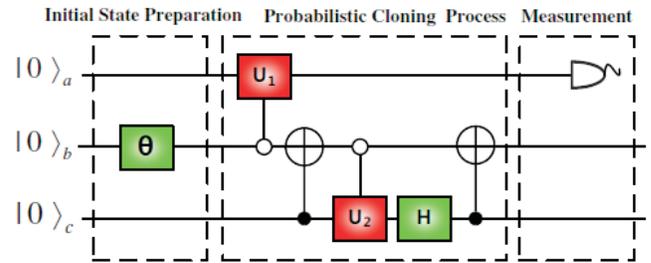}
 \caption{Quantum logic circuit for the
probabilistic quantum cloning demonstration. Qubit \emph{a} is the probe
qubit and qubit \emph{c} is the clone qubit, they are all initialized
in states $|0\rangle$. Qubit \emph{b} is initially prepared at the to-be-cloned qubit in
$|\psi_{in}\rangle_b=|\psi_{\pm\theta}\rangle$ through rotating qubit \emph{b} by angle $\pm\theta$
around the y-axis. The unitary
operation $U_{1}$ denotes $R_y(-\alpha)$ and $U_{2}$ denotes
$R_y(\beta)$, where
$\alpha=2\arccos(\sqrt{\frac{1+\tan^4\frac{\theta}{2}}{2}})$ and
$\beta=2\arccos((\sqrt{\frac{2}{1+\tan^4\frac{\theta}{2}}}+\sqrt{\frac{2}{1+\tan^{-4}\frac{\theta}{2}}})/2)$.
}\label{fig1}
\end{figure}

We start by introducing the experimental scheme to realize the
optimal $1\rightarrow2$ probabilistic quantum cloning. The quantum
logic circuit for the probabilistic quantum cloning process is
illustrated in Fig.~\ref{fig1}. This is a three-qubit network, which
contains one Hadamard gate, two controlled-NOT gates, and two
controlled-rotation gates.  In this network, qubit \emph{a} is the
probe qubit that indicates whether the cloning progress is
successful. Qubit \emph{b} is the to-be-cloned qubit, which is randomly chosen from
the set
$S=\{|\psi_{+\theta}\rangle,|\psi_{-\theta}\rangle\}$, where
\begin{equation}\label{input}
|\psi_{in}\rangle=|\psi_{\pm\theta}\rangle=\cos\frac{\theta}{2}|0\rangle\pm\sin\frac{\theta}{2}|1\rangle,\qquad
\theta\in[0,\frac{\pi}{2}].
\end{equation}
When $\theta=\pi/2$, the states in the set $S$ are orthogonal. This
simplification is reasonable because any pair of arbitrary states
$|\psi_1\rangle$ and $|\psi_2\rangle$ can be transformed to the form
of Eq.~(\ref{input}) via a unitary rotation. Qubit \emph{c} is the
cloning qubit and initially set to $|0\rangle$. After the unitary
evolution, if qubit \emph{a} is detected in state $|0\rangle$, the
cloning process succeeds and we will obtain two perfect copies of
the input state $|\psi_{in}\rangle$ at qubit \emph{b} and \emph{c};
and if qubit \emph{a} is detected in state $|1\rangle$, the cloning
process fails.

The output state at the end of the quantum circuit can be written as:
\begin{eqnarray}\label{unitary}
|\psi_{out}\rangle&=&U(|0\rangle|\psi_{in}\rangle|0\rangle) \nonumber\\
&=&\sqrt{\gamma}|0\rangle|\psi_{in}\rangle|\psi_{in}\rangle+\sqrt{1-\gamma}|1\rangle|\Phi\rangle_{BC},
\end{eqnarray}
where $U$ is a unitary operator for the quantum network shown in
Fig.~\ref{fig1}, the parameter $\gamma$ is the cloning efficiency
and
$|\Phi\rangle_{BC}=-\sqrt{1/(1+\tan^{4}\frac{\theta}{2})}|00\rangle_{BC}-\sqrt{1/(1+\tan^{-4}\frac{\theta}{2})}|11\rangle_{BC}$
is the normalized state of the composite system BC. The first term
denotes the success of the $1\rightarrow2$ cloning while the second
term represents the failure. If we substitute the expressions of the
unitary operator $U$ and the input states $|\psi_{in}\rangle$ in
Eq.~(\ref{unitary}), we obtain the cloning efficiency $\gamma$ as:
\begin{equation}\label{probability}
\gamma(\theta)=\frac{1}{1+|\langle\psi_{+\theta}|\psi_{-\theta}\rangle|}=\frac{1}{1+\cos\theta},
\end{equation}
which has been proven to be optimal \cite{Duan80,Pati83}. Compared
with the logic circuit of the cloning machine proposed in
Ref.\cite{Zhang61}, this scheme requires fewer quantum logic gates. This setup should be more robust in practice, as it is
less affected by experimental imperfections, such as errors in radio-frequency pulses and decoherence.

\begin{figure}[tbph]
  \centering
 \includegraphics[width=9 cm]{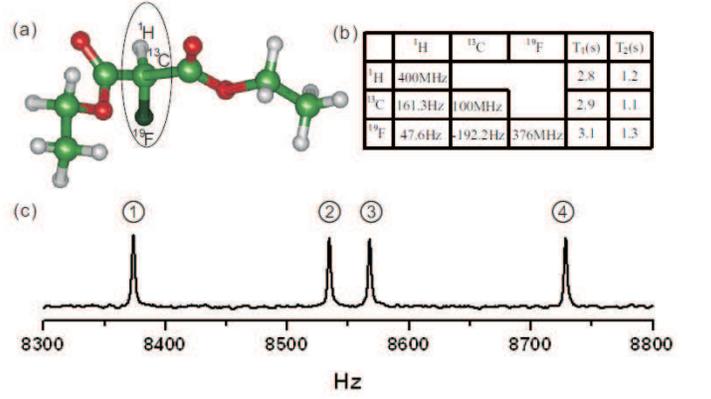}
  \caption{(color online) Molecular structure, NMR parameters and the $^{13}$C equilibrium spectrum of
Diethyl-fluoromalonate. (a) Molecular structure of
Diethyl-fluoromalonate. Three spin-$\frac{1}{2}$ nuclei as three
qubits are marked in oval. (b) The spin-spin couplings and chemical
shifts of the three nuclei. (c)The four main peaks in the spectrum
corresponding to signal read out from the $^{13}$C channel. They
respectively correspond to the states $|10\rangle$,
$|00\rangle$,$|11\rangle$ and $|01\rangle$ of $^1$H and $^{19}$F
from the left to the right. }\label{fig2}
\end{figure}

Using a sample of  Diethyl-fluoromalonate, the quantum circuit was
implemented on a liquid-state NMR quantum-information processor.
Three qubits are represented by the $^1$H, $^{13}$C and $^{19}$F
nuclear spins.  The molecular structure is shown in Fig.\ref{fig2}(a),
where the three nuclei used as qubits are marked by the oval. The
natural Hamiltonian of three-qubits system in the rotating frame can be written as:
\begin{equation}
H =\sum_{i=a}^{c} \omega_{i}I_{z}^{i}+2\pi\sum_{i<j}
J_{ij}I_{z}^{i}I_{z}^{j},\quad (i,j=a,b,c) \label{eq.H},
\end{equation}
where $\omega_{i}$ represent Larmor frequencies, $J_{ij}$'s are the
coupling constants: $J_{ab}=J_{HC}=161.3$ Hz, $J_{bc}=J_{CF}=-192.2$ Hz and
$J_{ac}=J_{HF}=47.6$ Hz.
The experiments were performed at room temperature using a Bruker
Avance  400MHz NMR spectrometer equipped with a QXI probe with pulsed
field gradient \cite{Zhang100,Peng103}.

The system was first prepared in a pseudo pure state (PPS)
$\rho_{000}=\epsilon|000\rangle\langle000|-\frac{1-\epsilon}{8}I$,
where $\epsilon\approx10^{-5}$ describes the thermal polarization of
the system and $I$ is the $8\times8$ identity matrix, using the
method of spatial averaging \cite{Cory120}. From the state
$\rho_{000}$, we prepared the initial state
$|0\rangle|\psi_{\pm\theta}\rangle|0\rangle$ through  rotating qubit
\emph{b} by angle $\pm\theta$ around the  $y$-axis.

\begin{figure}[tbph]
  \centering
\includegraphics[width=8.5 cm]{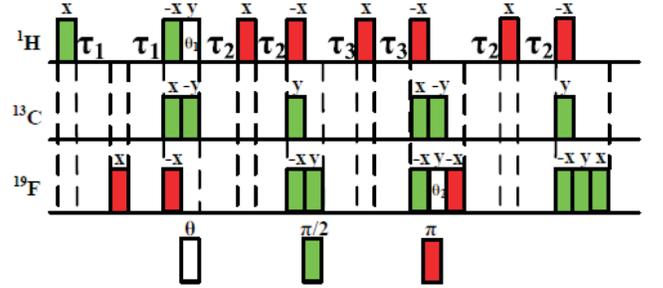}
\caption{Pulse sequence for the probabilistic quantum cloning process.
The circuit is implemented
by hard pulses and free evolutions. The grey rectangles denote
$\pi/2$ pulses and the black ones denote the refocusing $\pi$
pulses. The rectangles labeled with $\theta_i$ represent rotations
by an angle $\theta_i$, for $\theta_1=\alpha/2$ and
$\theta_2=(\beta+\pi)/2$. Pulse phases are shown up each pulse.
Delay times are $\tau_1=\frac{\alpha}{4\pi
J_{ab}}$,$\tau_2=\frac{1}{4J_{bc}}$,$\tau_3=\frac{\beta}{4\pi
J_{bc}}$.}\label{fig3}
\end{figure}

The quantum circuit is realized by hard pulses and free
evolution. The pulse sequence is depicted in Fig.~\ref{fig3}. In
principle, the readout procedure should be applied to each of the
cloning qubits in the subsequent experiments at the end of the quantum
circuit. In this experiment, a sample in natural abundance is used,
i.e., only $\approx 1\%$ of the molecules had one $^{13}$C nuclear
spin. To distinguish those molecules from the background molecules,
we collect all signals from the cloning qubits through the $^{13}$C channel, by applying
SWAP gates and measuring the $^{13}$C qubit. Figure \ref{fig2}(c)
shows the experimental spectra obtained by reading out the $^{13}$C
qubit. The spectrum consists of four resonance peaks, labeled by the
corresponding logical states of the corresponding logical states
$|10\rangle, |00\rangle, |11\rangle$ and $|01\rangle$ of nuclei
$^1$H and $^{19}$F. With respect to the status of probe qubit $^1$H,
we divide the four signal peaks into two groups: Group 1, including
peaks \textcircled{\small{1}} and \textcircled{\small{3}},
corresponds to state $|1\rangle$ of $^1$H, indicating the failure
cloning process; and group 2, $\ding {173}$ including peaks
\textcircled{\small{2}} and \textcircled{\small{4}}, corresponds to
state $|0\rangle$, indicating the success process. Unlike approximate quantum cloning, the probabilistic cloning machine will yield perfect cloning, and the faulty copies are rejected. In
the analysis, we filtered data from group 1, which represents the
failing case in the cloning process.

The signal intensity of the initial pseudopure state is
measured as the normalized reference. Then the relative signal
intensity is measured by digital quadrature detection (DQD). The
observable operator can be expressed as $\sigma_x+i\sigma_y$, so the
transverse components can be measured as
$P_x=\texttt{Tr}(\rho\sigma_x)$ and $P_y=\texttt{Tr}(i\rho\sigma_y)$
. In the subsequent experiment,  the vertical part $P_z$ is measured
by applying a $\pi/2$ pulse on the qubit after the cloning process.

It is well-established that probabilistic cloning machine are
analyzed with respect to two characteristics: cloning efficiency
$\gamma$ and cloning fidelity $F$. In the following, we will discuss
in details how we can obtain these two parameters from the spectrum
data. In the experiment, cloning efficiency $\gamma$ equals to the
population of the probe qubit $^1$H in the state $|0\rangle$. It can
be  measured by comparing the signal intensity of group 2 with the
total one, as shown in Eq.(\ref{efficiency:eq}).

\begin{eqnarray}\label{efficiency:eq}
P_i&=&|P_{2i}|+|P_{4i}|,\quad (i=x,y,z)\nonumber\\
\gamma&=&\sqrt{P^2_x+P^2_y+P^2_z}.
\end{eqnarray}

\begin{figure}[tbph]
  \centering
\includegraphics[width=8.5 cm]{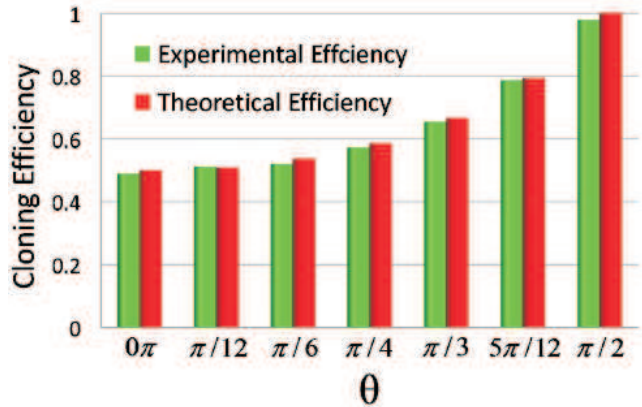}
\caption{(Color Online) Experimental efficiencies  versus different
angles $\theta$ of input state.  The average experimental
efficiencies for each $\theta$ are represented at green columns and
the red columns corresponding to the optimal values.}\label{fig4}
\end{figure}

We have studied the input states as the function of $\theta$, varied from
$0$ to $\pi/2$ in  $\pi/12$  increment. Fig.\ref{fig4} display the
average cloning efficiency $\gamma(\theta)$ (green columns) along
with the theoretical expectations (red columns). Note that the probabilistic quantum cloning machine
can produce faithful copies with probability $\gamma$.
The cloning efficiency is related to the distinguishable metric of
the quantum state space for the input states, since it increases
with decreasing of the overlap of the input states in the cloning
set $S$. The result clearly indicates that the larger the overlap
between the input states, the smaller the maximum cloning
efficiency. As the angle $\theta$ approaches $\frac{\pi}{2}$, the
cloning efficiency will be close to 1, which is possible if and only
if the states are chosen from an orthogonal set.

After the experimental cloning efficiencies were determined,
another important parameters, cloning fidelities, were analyzed. From the Bloch-sphere representation, the state of a
single qubit can be represented by a density matrix of the form
$\rho=(I+\vec{r}\cdot\vec{\sigma})$, where $I$ is the identity
operator and $\vec{\sigma}_\mu (\mu=x,y,z)$ are usual Pauli
matrices. Let
$\rho_0=|\psi_{in}\rangle\langle\psi_{in}|=\frac{1}{2}(I+\vec{r}_0\cdot\vec{\sigma})$
be the density matrix for the initial state, while
$\rho=\rho_b=\rho_c=\frac{1}{2}(I+\vec{r}_0\cdot\vec{\sigma})$ for
the copies. To experimentally determine the fidelities, we need the
density operators of the final states of both qubits. The length of
the vectors on the Block sphere representing the cloned states are
found to be  $r (\vec{r}_x, \vec{r}_y, \vec{r}_z)$, where
$r_i=(P_{2i}+P_{4i})/\gamma$ ($i=x,y,z$). Here, the signal
intensity should be divided by $\gamma$ to compensate the signal
loss of faulty copies. The cloning fidelities are expressed as
$$
F_{i}=\texttt{Tr}(\rho_{0}\cdot\rho_{i})=\frac{1}{2}(1+\vec{r}_{0}\cdot
\vec{r}_{i}),\,(i=b,c) ,
$$

For the initial state $|\psi_{in}\rangle =
\cos\frac{\theta}{2}|0\rangle\pm\sin\frac{\theta}{2}|1\rangle$, we
have $\vec{r}_0 = (\sin(\pm\theta), 0, \cos\theta)$ and the
fidelities become
\begin{equation}
F_i = \frac{1}{2} (1 +\sin(\pm\theta)\cdot r_{x_i} + \cos\theta\cdot
r_{z_i}) . \label{e.fid}
\end{equation}

\begin{figure}[tbph]
  \centering
\includegraphics[width=8.5 cm]{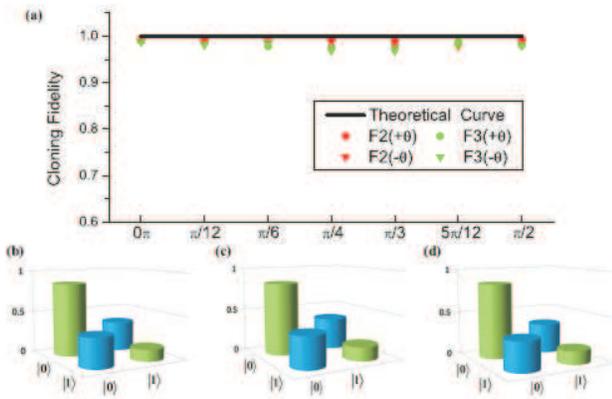}
\caption{(color online) Experimental  fidelities versus different
angles $\theta$ of input state. (a) The theoretical values of
cloning fidelity are plotted as the solid line. The cycles denote
the cloning fidelities for the second qubit (red) and third qubits
(green) for $|\psi(+\theta)\rangle$, while the triangles are for
$|\psi(-\theta)\rangle$.  (b)-(d) show the reconstructed density
matrices for $\theta=\pi/4$. (b) represents the matrix for the input
state $\rho_0$ and (c), (d) give the experimental results of
$\rho_b$ and $\rho_c$.}\label{fig5}
\end{figure}

The experimental fidelities of the duplicated states are shown in
Fig.\ref{fig5}. Fig.\ref{fig5}(b)-(d) show the reconstructed density
matrices for $\theta=\pi/4$. The vertical axes show the normalized
amplitude and the horizontal axes label the basis state in the
computational basis. Fig.\ref{fig5}(b) represents the matrix for the
input state $\rho_0$ and (c), (d) give the experimental results of
$\rho_b$ and $\rho_c$. The corresponding fidelities are $F_b=0.99$,
$F_c=0.98$. Fig.\ref{fig5}(a) shows the cloning fidelities for the
input states $|\psi(\pm\theta)\rangle$ with different angles
$\theta$ in the cloning set $S$. In the figure, the cycles denote
the cloning fidelities for the second qubit (red) and third qubits
(green) for $|\psi(+\theta)\rangle$, while the triangles are for
$|\psi(-\theta)\rangle$. The average experimental fidelity over
different values of $\theta$ is about $0.98$. The small deviations
($\leq3\%$) between the experimental and theoretical values are
mainly attributed to imperfect calibration of radio frequency
pulses. The decoherence from spin relaxation is negligible, since
the total experimental time of $\sim8$ ms is much shorter than the
minimal relaxation time of $\sim1.0$ s.

In summary, we have experimentally demonstrated the probabilistic
cloning machine in a NMR system by simplifying the network. For two
orthogonal states, the cloning machine can always produce two
perfect copies with the 100\% probability of success. When the
original states are non-orthogonal, two faithful copies can be
produced with the deterministic probability less than 1.  The
experimental results are in good agreement with the the theoretical
prediction, which indicates the cloning network is effective for the
all input states. Our experimental scheme can be achieved not only
in NMR systems, but also in other physical systems. Further
research works in probabilistic cloning, especially in
experimental demonstration, will be important for the other quantum
information protocols, such as quantum identification, quantum
purification and quantum deleting \cite{Duan80,Fiuravsek70,Feng65}.

We thank Prof. J. W. Pan and Dr.~W.~Harneit for helpful discussions. This work was
supported by the National Natural Science Foundation of China, the
National Science Foundation for Postdoctoral Scientists, the
Fundamental Research Funds for the Central Universities, the CAS,
Ministry of Education of PRC, and the National Fundamental Research
Program.



\begin{thebibliography}{99}
\bibitem{Wootters299}
W. K. Wootters and W. H. Zurek, Nature(London) \textbf{299},
802(1982); \ D. Dieks, Phys. Lett. \textbf{92A}, 271 (1982).
\bibitem{Buzek54}
V. Bu\v{z}ek,  and M. Hillery,  Phys. Rev. A  \textbf{54},
1844(1996).
\bibitem{Bennett84}
C. H. Bennett and G. Brassard,  in Proceedings of IEEE International
Conference on Computers, Systems and Signal Processing, Bangalore,
India (IEEE, New York, 1984), p. 175.
\bibitem{Niu60}
C.S. Niu and R. B. Griffiths, Phys. Rev. A \textbf{60}, 2764 (1999);
 \ N. J. Cerf et al., Phys. Rev. Lett. \textbf{88}, 127902 (2002); \ A.
Acin, N. Gisin, and V. Scarani, Phys. Rev. A \textbf{69}, 012309
(2004).
\bibitem{Scarani77}
V. Scarani, S. Iblisdir, N. Gisin and A. Ac$\acute{i}$n,  Rev. Mod.
Phys. \textbf{77}, 1225(2005).
\bibitem{Gisin79}
N. Gisin, and S. Massar,  Phys. Rev. Lett. \textbf{79}, 2153 (1997).
\bibitem{Cerf85}
N. J. Cerf, A. Ipe, and X. Rottenberg, Phys. Rev. Lett. \textbf{85},
1754 (2000).
\bibitem{Bruss62}
D. Bruss \emph{et al.,} Phys. Rev. A \textbf{62}, 012302 (2000).
\bibitem{Ariano64}
G. M. D’Ariano and P. Lo Presti, Phys. Rev. A \textbf{64}, 042308
(2001).

\bibitem{Delgado98}
Y. Delgado, L. Lamata, J. Le$\acute{o}$n, D. Salgado and E. Solano, Phys. Rev. Lett. \textbf{98},
150502(2007).
\bibitem{Lamata101}
L. Lamata, J. Le$\acute{o}$n, D. P$\acute{e}$rez-Garc$\acute{i}$a, D. Salgado and E. Solano, Phys. Rev. Lett. \textbf{101},
180506(2008).

\bibitem{Huang64}
Y. F. Huang,  \emph{et al.}, Phys. Rev. A \textbf{64}, 012315
(2001).
\bibitem{Antia296}
A. Lamas-Linares et al., Science \textbf{296}, 712 (2002).
\bibitem{Martini419}
F. De Martini, V. Bu\v{z}ek,  F. Sciarrino and  C. Sias, Nature
(London) \textbf{419}, 815 (2002).
\bibitem{Fasel89}
S. Fasel, N. Gisin, G. Ribordy, V. Scarani and H. Zbinden,  Phys.
Rev. Lett. \textbf{89}, 107901 (2002).
\bibitem{Zhao95}
Z. Zhao, \emph{et al.}, Phys. Rev. Lett. \textbf{95}, 030502 (2005).
\bibitem{Bartuskova99}
L. Bartuskova,\emph{et al.}, Phys. Rev. Lett. \textbf{99},
120505(2007).

\bibitem{Nagli720}
E. Nagali, L. Sansoni, F. Sciarrino, \emph{et al.}, NATURE PHOTONICS
\textbf{3}, 720(2009).

\bibitem{Cummins88}
 H. K. Cummins, \emph{et al.} Phys. Rev. Lett. \textbf{88}, 187901
(2002).
\bibitem{Du94}
J. F. Du,  \emph{et al.}, Phys. Rev. Lett. \textbf{94}, 040505
(2005).
\bibitem{Chen75}
H. W. Chen, X. Y. Zhou, Dieter Suter and J. F. Du,   Phys. Rev. A
\textbf{75}, 012317 (2007).
\bibitem{Duan80}
L. M. Duan and G. C. Guo,  Phys. Rev. Lett. \textbf{80}, 4999(1998);
\ L. M. Duan and G. C. Guo,  Phys. Lett. A \textbf{243}, 261(1998).
\bibitem{Pati83}
A. K. Pati, Phys. Rev. Lett. \textbf{83}, 2849(1999).
\bibitem{Qiu39}
D. W. Qiu, J. Phys. A: Math. Gen. \textbf{39}, 5135(2006).
\bibitem{Azuma72}
K. Azuma, Phys. Rev. A. \textbf{72}, 032335 (2005).
\bibitem{Zhang61}
C. W. Zhang, Z. Y. Wang, C. F. Li and G. C. Guo,  Phys. Rev. A
\textbf{61}, 062310(2000);\quad T. Gao \emph{et al}., e-print
arXiv:quant-ph/0308036.
\bibitem{Zhang100}
 J. F. Zhang, X. H. Peng, \emph{et al.} Phys. Rev. Lett. \textbf{100},
 100501(2008).
\bibitem{Peng103}
X. H. Peng, J. F. Zhang\emph{et al.} Phys. Rev. Lett. \textbf{103},
 140501(2009).
\bibitem{Cory120}
D. G. Cory, M. D. Price and T. F. Havel, Phys. D \textbf{120},
82(1998).
\bibitem{Fiuravsek70}
J. Fiur$\acute{a}\check{s}$ek,  Phys. Rev. A. \textbf{70}, 032308
(2004).
\bibitem{Feng65}
Y. Feng, S. Y. Zhang, and M. S. Ying,  Phys. Rev. A. \textbf{65},
042324 (2002).
\end{thebibliography}
\end{document}